\documentstyle[12pt,epsf]{article}

\newcommand{\idxx}[2]{\frac{d^{#1} #2}{(2\pi)^{#1}i}}

\newcommand{\tr}{\mbox{tr}}

\def\fsl#1{\setbox0=\hbox{$#1$}           
   \dimen0=\wd0                                 
   \setbox1=\hbox{/} \dimen1=\wd1               
   \ifdim\dimen0>\dimen1                        
      \rlap{\hbox to \dimen0{\hfil/\hfil}}      
      #1                                        
   \else                                        
      \rlap{\hbox to \dimen1{\hfil$#1$\hfil}}   
      /                                         
   \fi}                                         %

\begin{document}

\begin{flushright}
\begin{minipage}[t]{4cm}
DPNU-98-17 \\
hep-th/9805035 \\
May 1998
\end{minipage}
\end{flushright}

\vspace{0.3cm}

\begin{center}

{\LARGE The Higgs Boson Mass and Ward-Takahashi Identity \\
in Gauged Nambu-Jona-Lasinio Model}

\hspace{1cm}

{\large Michio Hashimoto
\vspace{2mm}\\
{\it Department of Physics, Nagoya University}
\vspace{2mm}\\
{\it Nagoya 464-8602, Japan}}

\end{center}

\baselineskip=.285in

\vspace{1cm}

\begin{abstract}
A new formula for the composite Higgs boson mass is given, 
based on the Ward-Takahashi identity and the Schwinger-Dyson(SD) equation. 
In this formula the dominant asymptotic solution of the SD equation 
yields a correct answer, in sharp contrast to 
the Partially Conserved Dilatation Current(PCDC) approach where 
the sub- and sub-sub-dominant solutions should be taken into account 
carefully. 
In the gauged Nambu-Jona-Lasinio model we find $M_H \simeq \sqrt{2}M$ 
for the composite Higgs boson mass $M_H$ and the dynamical mass of 
the fermion $M$ in the case of the constant gauge coupling(with large 
cut off), which is consistent with the PCDC approach 
and the renormalization-group approach. 
As to the case of the running gauge coupling, we find 
$M_H \simeq 2 \sqrt{(A-1)/(2A-1)}M$, 
where $A \equiv 18 C_2 /(11N_c - 2N_f)$ with 
$C_2$ being the quadratic Casimir of the fermion representation. 
We also discuss a straightforward application of our formula 
to QCD(without 4-Fermi coupling), which yields 
$M_{\sigma} \sim \sqrt{2}M_{dyn}$, 
with $M_{\sigma}$ and $M_{dyn}$ being the light 
scalar(``$\sigma$-meson'') mass and mass of the constituent quark, 
respectively. 
\end{abstract}

The Nambu-Jona-Lasinio(NJL) model \cite{NJL}, which is not renormalizable 
in 4 dimensions, is familiar to us as an example of model 
for the dynamical symmetry breaking. 
The Gauged Nambu-Jona-Lasinio(GNJL) model 
has been studied vigorously \cite{GNJL,KMY} and 
is known to be {\it renormalizable} in 4 dimensions \cite{RGNJL}. 
Phenomenologically, 
it has been applied to the Top Mode Standard Model(TMSM) \cite{MTY}. 
Unfortunately, however, it has been complicated to calculate 
the composite Higgs boson mass in this model by the approach based on 
the Schwinger-Dyson(SD) equation. 
Namely, the previous manner to estimate the composite Higgs boson 
mass \cite{STY} 
was based on the Partially Conserved Dilatation Current (PCDC) relation 
\cite{PCDC}, where we needed inevitably to know 
the vacuum energy in the broken phase by way of 
the Cornwall-Jackiw-Tomboulis(CJT) potential \cite{CJT}. 
The PCDC relation is \cite{PCDC} 
\begin{equation}
 M_H^2 = - 16 \frac{d_{\theta}E}{d_{\sigma}^2 f_{\pi}^2} , \label{24}
\end{equation}
where $E$, $d_{\sigma}$ and $d_{\theta}$ are the vacuum energy, 
the scale dimension of $\sigma$ and that of the trace of the energy-momentum 
tensor, respectively. In the GNJL model, we know 
$d_{\theta}=2d_{\sigma}$ and 
$d_{\sigma} = 2-\sqrt{1-\lambda/\lambda_c}$ \cite{KSY}, where 
$\lambda \equiv 3 C_2 \alpha /4\pi $ and $ \lambda_c = 1/4$, 
with $g$ and $C_2$ being the gauge coupling constant and 
the quadratic Casimir of the fundamental representation, respectively. 

In this manner, the result based on the full non-linear SD equation 
yields $M_H \simeq \sqrt{2}M$ \cite{STY,cw}, 
while the linearized solution does $M_H \simeq 2M$ for 
$\lambda \ll 1$ \cite{STY}. 
It implies that the usual bifurcation method \cite{bifurcation} 
does not work in this method. 
Besides the SD approach, on the other hand, 
we can obtain the same result as that of the full non-linear one, 
$M_H \simeq \sqrt{2}M$, 
by the renormalization-group(RG) approach \cite{Higgsmass_RGE,cw}. 
In the RG approach, of course, we need not know the vacuum energy in 
the broken phase. 
Thus, it would be natural to seek another method without 
using the vacuum energy even in the SD approach. 

In this paper, we derive a formula for the composite Higgs boson mass 
based on the Ward-Takahashi (WT) identity by using the technique of 
Ref.~\cite{Higgsmass_WT}. 
Then, we obtain {\it analytically} the same result as the RG approach 
in the case of the constant gauge coupling and new result for 
the running gauge coupling. 
Finally, we simply apply our method to the light scalar meson
(``$\sigma$ meson'') in the QCD. 

Let us consider the $ SU(N_c) $-gauged NJL model, 
with $ U(1)_L \times U(1)_R $ chiral symmetry for simplicity: 
\begin{eqnarray}
{\cal L} & = & \bar{\psi} ( i \fsl{\partial} - g \fsl{A} ) \psi 
+ \frac{G}{2N_c} \left[ ( \bar{\psi} \psi )^2 + 
( \bar{\psi} i \gamma_5 \psi )^2 \right]
- \frac{1}{2} \tr ( F_{\mu\nu} F^{\mu\nu} ) \\
& \to & \bar{\psi} ( i \fsl{\partial} - g \fsl{A} ) \psi 
-\bar{\psi} ( \sigma + i \gamma_5 \pi ) \psi 
- \frac{N_c}{2 G} ( \sigma^2 + \pi^2 ) 
- \frac{1}{2} \tr ( F_{\mu\nu} F^{\mu\nu} ) , \label{lag}
\end{eqnarray}
where we have used the auxiliary field method, 
$ \sigma \propto \bar{\psi}\psi $ and 
$ \pi \propto \bar{\psi} i \gamma_5 \psi $, 
and $ \psi $ belongs to the fundamental representation of $ SU(N_c) $, 
and $ g $ and $ G $ are the gauge coupling and the 4-Fermi coupling, 
respectively. 
Following Ref.~\cite{Higgsmass_WT}, we consider 
the partition function including the source terms for the composite 
bosons from the beginning, which yields desired WT identities. 
Let us consider the following partition function:
\begin{eqnarray}
 Z[\eta, \bar{\eta}, J_{A_{\mu}}, J_{\sigma}, J_{\pi}] & \equiv & 
\int {\cal D}\psi {\cal D}\bar{\psi} {\cal D}A_{\mu} {\cal D}\sigma 
{\cal D} \pi \exp \left( i \int dx^4 {\cal L} + {\cal L}_{sourse} \right), \\
{\cal L}_{sourse} & \equiv & \bar{\psi} \eta + \bar{\eta} \psi + 
                             J_{A_{\mu}} A_{\mu} + J_{\sigma}\sigma + 
                             J_{\pi}\pi . 
\end{eqnarray} 
Then, we easily obtain the chiral WT identity as follows:
\begin{equation}
 \partial_{\mu} \langle \bar{\psi}\gamma^{\mu}\gamma_5 \psi \rangle_J 
  + i \langle \bar{\eta}\gamma_5 \psi \rangle_J 
  + i \langle \bar{\psi}\gamma_5 \eta \rangle_J  
  + 2i \langle J_{\sigma} \pi \rangle_J 
  - 2i \langle J_{\pi} \sigma \rangle_J = 0 . \label{WT}
\end{equation}
In the usual way, we define the effective action 
$\Gamma [\bar{\psi}, \psi, A_{\mu}, \sigma, \pi]$ by 
the Legendre transformation of $W[J]$, where $W[J]$ is the generating 
functional for the connected Green function~$(iW[J] \equiv \ln Z[J])$. 
After we rewrite Eq.~(\ref{WT}) in terms of 
$\Gamma [\bar{\psi}, \psi, A_{\mu}, \sigma, \pi]$, $\psi$, $\bar{\psi}$, 
$\sigma$ and $\pi$, we take the derivative with respect to 
$\psi$ and $\bar{\psi}$. 
By taking the Fourier transformation, we obtain 
the WT identity for the axial-vector vertex as follows:
\begin{eqnarray}
 q_{\mu}\Gamma_5^{\mu} (p+q,p) & = & i S_f^{-1}(p+q) \gamma_5 + 
                                 i \gamma_5 S_f^{-1}(p) \nonumber \\
& &  \hspace{-1cm} 
     -2 \frac{\delta^3 \Gamma [\bar{\psi}, \psi, A_{\mu}, \sigma, \pi]}
      {\delta \psi \delta \bar{\psi} \delta \sigma}(p+q,p) \pi (q) 
     +2 \frac{\delta^3 \Gamma [\bar{\psi}, \psi, A_{\mu}, \sigma, \pi]}
      {\delta \psi \delta \bar{\psi} \delta \pi}(p+q,p) \sigma (q) . 
                                                               \label{7}
\end{eqnarray}
On the other hand, by definition of $f_{\pi}$, 
the decay constant of NG boson, it is well-known \cite{DSBbook} 
that the following relation holds in the limit of $q^{\mu} \to 0$ : 
\begin{equation}
 q_{\mu}\Gamma_5^{\mu} (p+q,p) = i S_f^{-1}(p+q) \gamma_5 + 
                                 i \gamma_5 S_f^{-1}(p) 
  +\chi_{\pi} (p,q) f_{\pi} \qquad (q^{\mu} \to 0) , \label{8}
\end{equation}
where $\Gamma_5^{\mu}(p+q,p)$ is the proper axial-vector vertex 
and $\chi_{\pi}(p,q) $ is the amputated Bethe-Salpeter(BS) amplitude 
defined by $\chi_{\pi}(p,q) \equiv S_f^{-1}(p+q) {\cal F. T. } 
\langle 0|\mbox{T}\psi (x) \bar{\psi}(0) | \pi(q) \rangle S_f^{-1}(p)$. 
Comparing Eq.~(\ref{7}) with Eq.~(\ref{8}), 
we obtain 
\begin{eqnarray}
 f_{\pi}&=& 2 Z^{-1/2}_{\pi}(0) \langle \sigma \rangle , \label{9} \\
 \chi_{\pi}(p,0) &=& 2 \gamma_5 \frac{\Sigma (p)}{f_{\pi}} ,
\end{eqnarray}
where we used $iS_f^{-1}(p) \equiv \fsl{p} - \Sigma (p) $ and 
$ \langle \pi \rangle =0$ and 
$\pi = Z_{\pi}^{1/2}\pi_R$. 
Notice that in Eq.~(\ref{7}) the proper axial-vector vertex is directly 
obtained by the Legendre transformation. 

In addition to these relations, we can derive the WT identity 
for the composite boson propagator by differentiating 
Eq.~(\ref{WT}) with respect to $J_{\sigma}$ and $J_{\pi}$ :
\begin{eqnarray}
 \partial_{\mu}^x \langle \bar{\psi}(x) \gamma^{\mu}\gamma_5 \psi(x) 
    \sigma (y) \pi (z) \rangle_{J=0} 
   + 2i \langle \pi (x) \pi (z) \rangle_{J=0} \delta (x-y)
  - 2i \langle \sigma (x) \sigma (y) \rangle_{J = 0} \delta (x-z)=0 . 
                                    \label{WT_Higgs}
\end{eqnarray}
After taking the Fourier transformation of (\ref{WT_Higgs}), 
multiplying it by 
$\frac{\delta^2 \Gamma [\bar{\psi},\psi,A_{\mu},\sigma,\pi]}
      {\delta \sigma \delta \sigma}$ and 
$\frac{\delta^2 \Gamma [\bar{\psi},\psi,A_{\mu},\sigma,\pi]}
      {\delta \pi \delta \pi}$ 
and taking $ q^{\mu} \to 0$ limit, we obtain the composite Higgs boson 
mass as 
\begin{eqnarray}
 Z_{\sigma}^{-1}(0) M_H^2 (0) & = & -\frac{Z_{\pi}^{1/2}(0) f_{\pi}}{2} 
        \frac{\delta^3 \Gamma [\bar{\psi},\psi,A_{\mu},\sigma,\pi]}
            {\delta \sigma \delta \pi \delta \pi}(0,0,0) \\
 & = & -\langle \sigma \rangle \Gamma_{\sigma,\pi,\pi}(0,0,0) , \label{14}
\end{eqnarray}
where we used Eq.~(\ref{9}), and $Z_{\sigma}^{-1}(p)M_H^2 (p)$ is defined 
through the $\sigma$-propagator 
$D_{\sigma}^{-1}(p) = 
\frac{\delta^2 \Gamma [\bar{\psi},\psi,A_{\mu},\sigma,\pi]}
     {\delta \sigma \delta \sigma}(p) \equiv 
Z_{\sigma}^{-1}(p) (p^2 - M_H^2(p)) $ and 
$\Gamma_{\sigma, \pi ,\pi}(0,0,0) \equiv \frac{\delta^3 \Gamma 
[\bar{\psi},\psi,A_{\mu},\sigma,\pi]}
{\delta \sigma \delta \pi \delta \pi}(0,0,0) $. 
Since Eq.~(\ref{14}) is derived through WT identity alone, 
Eq.~(\ref{14}) holds generally even in other models. 
In fact, we can easily confirm it in the linear-$\sigma$ model 
or the Standard Model(SM) at tree level. 
Thus, we may consider Eq.~(\ref{14}) as a generalization 
of the tree level relation in these models. 

%
%
\begin{figure}[t]
  \begin{center}
    \leavevmode
     \epsfxsize= 15cm \epsfbox{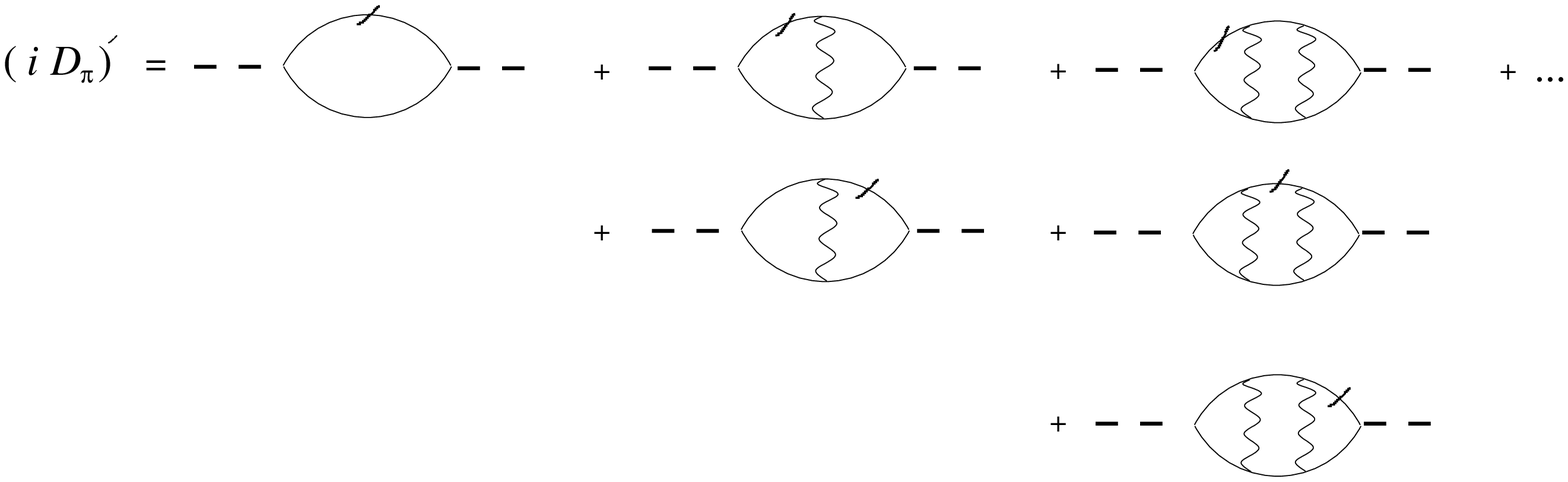}%
  \end{center}
\caption{{\scriptsize Insertion of $\sigma$ into the fermion propagator 
in the expression of the composite boson propagator $D_{\pi}^{-1} (0)$. 
The solid line, the doted external line, the wave line and 
the prime sign represent the full fermion propagator $ S_f $, 
the composite boson $\pi$, the gluon propagator and 
insertion of $\sigma$ at zero momentum into the fermion propagator, 
respectively. 
Notice that insertion of $\sigma$ into the gluon propagator 
is in higher order of the $1/N_c$-expansion. 
}}
\label{fig1}
\end{figure}
Now we come to evaluation of $\Gamma_{\sigma,\pi ,\pi}(0,0,0)$. 
It is rather difficult to obtain {\it non-perturbatively} the 3-point vertex 
$\Gamma_{\sigma,\pi ,\pi}(0,0,0)$ in general case. 
Fortunately, in the GNJL model we can easily estimate this 3-point vertex 
by using the {\it non-perturbative} propagator of $\pi$. 
The propagator of the composite boson in the GNJL model was obtained by 
Appelquist, Terning and Wijewardhana \cite{ATW}, whose technique was 
based on the resummation of the Taylor series around zero momentum of 
the composite boson under certain approximations. 
Recently, Gusynin and Reenders have given analytically 
the composite boson propagator in other approach 
without using the resummation \cite{Gusynin}. 
As far as we discuss $\Gamma_{\sigma,\pi,\pi}(0,0,0)$, however, 
the resummation technique is enough. 
We obtain the 3-point vertex $\Gamma_{\sigma,\pi,\pi}(0,0,0)$ by 
insertion of $\sigma$ with zero momentum into the fermion propagator 
in the expression of $iD_{\pi}^{-1}(0)$, 
which consists of the ladder graphs, (see Fig.~1). 
Notice that insertion of $\sigma$ into the gluon propagator is similar to 
the vacuum polarization graph and is in higher-order of 
the $1/N_c$-expansion. 
By using the resummation technique, we find that 
$\Gamma_{\sigma,\pi,\pi}(0,0,0)$ is graphically equal to Fig.~2 at 
$1/N_c$-leading order. 
%
%
\begin{figure}[t]
  \begin{center}
    \leavevmode
     \epsfxsize= 15cm \epsfysize=7cm \epsfbox{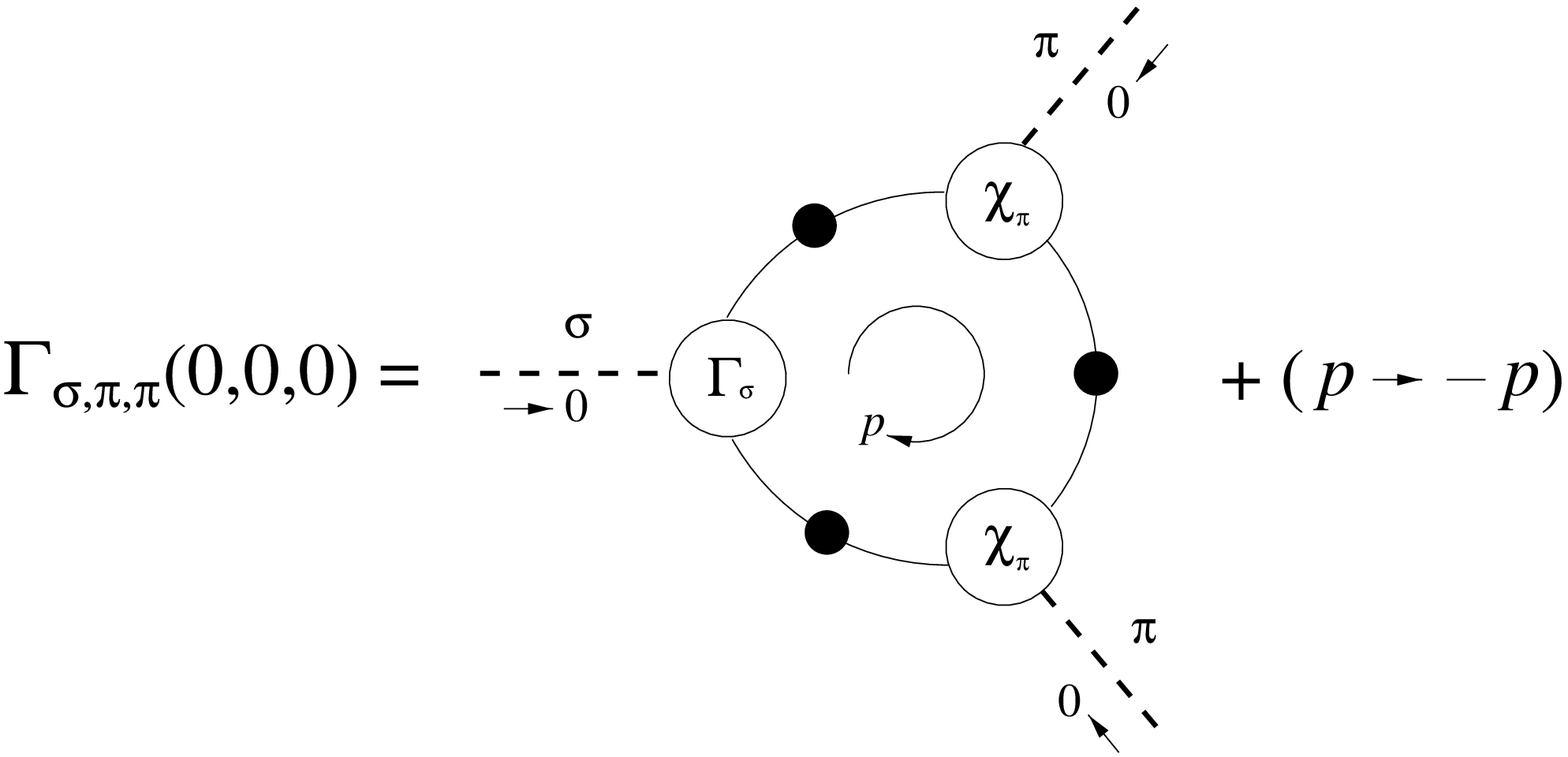}%
  \end{center}
\caption{{\scriptsize The 3-point vertex $\Gamma_{\sigma,\pi,\pi}(0,0,0)$. 
The solid line with shaded blob and the doted external line represent 
the full fermion propagator $ S_f $ and 
the composite bosons $\sigma,\pi$, respectively, 
while $\Gamma_{\sigma}$ and $\chi_{\pi}$ represent 
the $\sigma$-vertex 
($ \Gamma_{\sigma}(p)=\frac{d \Sigma(p)}{d \sigma}$~\cite{RGNJL}~) 
and the $\pi$-vertex at zero momentum, respectively. 
}}
\label{fig2}
\end{figure}
Thus, $\Gamma_{\sigma,\pi,\pi}(0,0,0)$ is written by 
\begin{eqnarray}
 \Gamma_{\sigma,\pi,\pi}(0,0,0) 
 & = & -2 N_c Z_{\pi}^{-1}(0) \times \nonumber \\
 & &    \int \idxx{4}{p} \tr \left[ 
          \frac{1}{\fsl{p}-\Sigma (p) } \frac{2\gamma_5 \Sigma(p)}{f_{\pi}}
          \frac{1}{\fsl{p}-\Sigma (p) } \frac{2\gamma_5 \Sigma(p)}{f_{\pi}}
          \frac{1}{\fsl{p}-\Sigma (p) } \frac{d \Sigma (p)}{d \sigma}
\right]  \\
 & = & -2 N_c Z_{\pi}^{-1}(0) \frac{1}{\pi^2}\int dx x  
          \frac{\Sigma^3 (x) \frac{d \Sigma(x)}{d \sigma}}
               {f_{\pi}^2(x+\Sigma^2(x))^2} \qquad (x=-p^2) .
\label{17}
\end{eqnarray}
As is well-known, $f_{\pi}$ is estimated by the Pagels-Stokar(PS) 
formula \cite{PS}:
\begin{equation}
 f_{\pi}^2 = \frac{N_c}{2\pi^2} \int dx x 
              \frac{\Sigma^2 (x)-\frac{x}{4}\frac{d \Sigma^2(x)}{dx}}
                   {(x+\Sigma^2(x))^2} . \label{18}
\end{equation}
Combining Eq.~(\ref{14}) with Eqs.~(\ref{17}) and (\ref{18}), 
we finally obtain the formula for the composite Higgs boson mass 
as follows:
\begin{equation}
 \frac{M_H^2}{4}  =\int dx x  
   \frac{\Sigma^3 (x) \frac{d \Sigma(x)}{d \sigma} \langle \sigma \rangle}
        {(x+\Sigma^2(x))^2} \left/ 
                   \int dx x 
    \frac{\Sigma^2 (x)-\frac{x}{4}\frac{d \Sigma^2(x)}{dx}}
         {(x+\Sigma^2(x))^2} \right.  , \label{Higgsmass1}
\end{equation}
where we used $Z_{\pi}(0) \simeq Z_{\sigma}(0)$ up to 
${\cal O}(M/\Lambda)$ due to the chiral symmetry 
and defined $M_H \equiv M_H(0)$. 
Of course, we can easily extend this result to 
the $SU(N_f)_L \times SU(N_f)_R$-symmetric GNJL model or TMSM. 
Note that Eq.~(\ref{Higgsmass1}) reproduces the well-known NJL result 
$M_H^2 = 4\sigma^2$ in the pure NJL limit($\Sigma(x) = \sigma$). 

In the case of the constant gauge coupling, 
we know the solution of the SD equation with one gluon exchange 
graph as \cite{KMY}
\begin{eqnarray}
 \Sigma (x) & \simeq & M \left( \frac{x}{M^2} 
     \right)^{\frac{-1+\sqrt{1-\lambda/\lambda_c}}{2}} , \label{21} \\
 \sigma & = & \Sigma (\Lambda^2) + \Lambda^2 \Sigma' (\Lambda^2) , \label{22}
\end{eqnarray}
where the bifurcation method \cite{bifurcation} is used and 
the infrared mass $M$ is defined by $\Sigma(M^2) = M$. 
From Eqs.~(\ref{21}) and (\ref{22}) the composite Higgs boson mass is 
obtained as
\begin{equation}
 \frac{M_H^2}{2M^2} \simeq  1 + \left( \frac{M^2}{\Lambda^2} 
              \right)^{1-\sqrt{1-\lambda/\lambda_c}} , \label{23}
\end{equation}
where we used $\frac{d \Sigma(x)}{d \sigma} \simeq 
(x/\Lambda^2)^{\frac{-1+\sqrt{1-\lambda/\lambda_c}}{2}}$ \cite{RGNJL} and 
assumed $M_H^2(0) \simeq M_H^2(M^2) \simeq M_H^2(M_H^2)$. 
Eq.~(\ref{23}) yields $M_H \simeq \sqrt{2} M$ in the limit of 
$ \Lambda \to \infty (\lambda \ne 0)$, which is the same result 
as that obtained by Shuto, Tanabashi and Yamawaki through 
the PCDC relation \cite{STY}.  
Eq.~(\ref{23}) also agrees with the RG analysis 
\cite{Higgsmass_RGE,cw} for a small gauge coupling 
$(1-\sqrt{1-\lambda/\lambda_c} \simeq \frac{\lambda}{2\lambda_c})$. 
In Ref.~\cite{cw}, Carena and Wagner also obtained 
the same result as Eq.~(\ref{23}) 
by estimating the curvature of the effective potential in terms of 
the solution of the SD equation. 
However, both methods \cite{STY,cw} essentially depend on 
the sub- and sub-sub-dominant solutions in the asymptotic expansion 
of the SD equation. 
The coefficient of the dominant term is obtained uniquely 
in various linearized methods, like the bifurcation method, 
the manner of replacing the mass function in the denominator of 
the SD equation by a constant mass, i.e., ($x+\Sigma^2(x) \to x+m^2$) 
and the power expansion for the SD equation, etc.. 
On the other hand, coefficients of the sub-, or sub-sub-dominant 
term are quite different depending on the linearized methods. 
(Compare Ref.~\cite{RGNJL} with Ref.~\cite{cw}). 
On the contrary, in our method the dominant solution of the SD equation 
already gives the desirable answer 
without using the sub- and sub-sub-dominant terms, and hence 
our formula (\ref{Higgsmass1}) is 
more convenient than the previous ones. 

In the case of the running coupling, the solution of the SD equation is 
\cite{KSY}
\begin{equation}
 \Sigma (x) \simeq M \left( \frac{\ln x/ \Lambda_{QCD}^2}
                                 {\ln M^2/\Lambda_{QCD}^2}\right)^{-A/2},
\end{equation}
where we used the usual improved ladder calculation \cite{higashijima} 
and $\sigma \simeq \Sigma(\Lambda^2)$ and $ A \equiv 18C_2/(11N_c -2N_f)$. 
Thus, we can obtain the composite Higgs mass in the GNJL model as follows:
\begin{equation}
 \frac{M_H^2}{2M^2} \simeq \frac{2(A-1)}{2A-1}
                   \left( \ln M^2/\Lambda_{QCD}^2\right)^{A}
                   \frac{\left( \ln M^2/\Lambda_{QCD}^2\right)^{-2A+1}-
                          \left( \ln \Lambda^2/\Lambda_{QCD}^2\right)^{-2A+1}}
                        {\left( \ln M^2/\Lambda_{QCD}^2\right)^{-A+1}-
                          \left( \ln \Lambda^2/\Lambda_{QCD}^2\right)^{-A+1}}.
            \label{25}
\end{equation}
This result is also consistent with RGE analysis 
at $1/N_c$-leading order. 
In the limit of $\Lambda \to \infty$, Eq.~(\ref{25}) gives 
$M_H = 2\sqrt{(A-1)/(2A-1)}M$, which coincides with 
$M_H \simeq \sqrt{2}M$ of Eq.~(\ref{23}) in the limit of 
$A \to \infty$ (non-running limit). 

Even if we consider pure QCD, i.e., the pure gauge limit of 
the GNJL model(without 4-Fermi coupling), 
we can obtain the same relation as Eq.~(\ref{Higgsmass1}) by 
repeating the previous derivation of the WT identity for 
the bi-local fields $\sigma \propto \bar{\psi}\psi $ and 
$\pi^a \propto \bar{\psi} i \gamma_5 T^a \psi$. 
The solution of the SD equation is obtained as follows \cite{DSBbook}:
\begin{equation}
 \Sigma (x) \simeq \frac{C}{x} 
             \left( \ln x/ \Lambda_{QCD}^2 \right)^{A/2-1} 
\qquad (x \gg \Lambda_{QCD}^2) ,
\end{equation}
where the coefficient of $C$ is equal to 
$M_{dyn}\mu^2 (\ln \mu^2/\Lambda_{QCD}^2)^{-A/2+1}$ 
under the condition of $M_{dyn}=\Sigma (\mu^2)$ 
at the renormalization point $\mu$ and 
$M_{dyn}$ is the constituent quark mass $(M_{dyn} \sim 320 \mbox{MeV})$. 
Now, we divide the integrations in Eq.~(\ref{Higgsmass1}) into 
the infrared part$(\int_0^{\mu^2})$ and 
the asymptotic one$(\int_{\mu^2}^{\Lambda^2})$ and define 
$ I_{infra}+I $ for the integration of the numerator 
and $J_{infra}+J$ for that of the denominator, 
where $I/M_{dyn}^4$ and $J/M_{dyn}^2$ are numerically equal to 
0.22 and 0.44 for $\mu = $ 1 GeV and $\Lambda_{QCD}\sim $ 200 MeV, 
respectively. 
%
%
%
%
Thus, we find that $J_{infra}$ gives the same order contributions 
as that of $J$ to the decay constant $f_{\pi}$ 
in contrast to the case of the GNJL model with a small gauge coupling. 
It is very difficult to estimate the infrared correction of $J_{infra}$ 
without assumptions. We simply assume that 
the mass function depends linearly on $x$ in the infrared region as 
$
 \Sigma (x) \simeq \frac{M_{dyn}-\Sigma(0)}{\mu^2} x + \Sigma(0) \quad
 (0 \leq x \leq \mu^2).
$
We find $\Sigma (0) \sim $ 100 MeV, provided that the PS formula including 
the correction of $J_{infra}$ gives $f_{\pi} = 93$ MeV. 
Then, we obtain $M_{\sigma}^2 \sim 2.1 \times M_{dyn}^2$ 
by using Eq.~(\ref{Higgsmass1}). This result is not changed for 
$N_f = 2 \sim 5$. 

\vspace{0.5cm}

In summary, a new formula for the composite Higgs boson 
mass~(\ref{Higgsmass1}) was derived from the WT identity. 
In the GNJL model, the dominant solution of the SD equation gives 
the desirable answer consistent with the previous 
calculations \cite{STY,Higgsmass_RGE,cw}. 
Thus, our method is more convenient 
than the manner based on the PCDC relation which depends on 
the sub- and sub-sub-dominant terms. 
By using the formula of Eq.~(\ref{Higgsmass1}), the same analytical result 
(\ref{23}) as that from the RGE analysis was obtained easily 
in the case of the constant gauge 
coupling. If the running coupling effects are taken into account, 
the new relation of $M_H \simeq 2\sqrt{(A-1)/(2A-1)}M$ is obtained 
for the GNJL model in the SD approach. 

Our approach may also be applied to QCD. 
By using our formula naively, we obtain the scalar meson mass as 
$M_{\sigma} \sim \sqrt{2}M_{dyn}$, i.e., around 500 MeV, 
where the decay constant is estimated by the PS formula. 
It is consistent with the light scalar $\sigma$ meson, whose mass 
is 400-1200 MeV in Particle Data Group \cite{pdg} 
and is recently estimated to be 400-600 MeV \cite{sigmameson}. 
Of course, $f_{\pi}$ may be estimated by other approach instead of 
the PS formula, for instance, by the ladder exact approach \cite{kugomich}, 
which, however, would not change significantly the relation of 
$M_{\sigma} \sim \sqrt{2}M_{dyn}$ in QCD. 

A problem would be to clarify the relation between our formula and 
the PCDC relation, both of which yield similar results at least numerically. 
Actually, our Higgs mass formula of Eq.~(\ref{Higgsmass1}) does not 
look like a simple modification of the PCDC relation of Eq.~(\ref{24}). 
This will be studied in future work. 

\vspace{0.5cm}

The author is very grateful to K.~Yamawaki for 
helpful discussions and reading this manuscript carefully. 
Thanks are also due to M.~Sugiura and V.~A.~Miransky for discussions.


\begin{thebibliography}{99}
 \bibitem{NJL} Y.~Nambu and G.~Jona-Lasinio, Phys. Rev. {\bf 122} (1961) 
	 345. 

 \bibitem{GNJL} W.~A.~Bardeen, C.~N.~Leung and S.~T.~ Love, 
  Phys. Rev. Lett. {\bf 56} (1986) 1230; 
  C.~N.~Leung, S.~T.~Love and  W.~A.~Bardeen, Nucl. Phys. {\bf B273} 
  (1986) 649.   

 \bibitem{KMY} K.-I.~Kondo, H.~Mino, and K.~Yamawaki, 
  Phys. Rev. {\bf D39} (1989) 2430; 
  K.~Yamawaki, 
  in {\it Proc. Johns Hopkins Workshop on Current Problems in Particle 
  Theory 12, Baltimore, June 8-10, 1988}, 
  edited by G. Domokos and S. Kovesi-Domokos 
  (World Scientific Pub. Co. , Singapore 1988); 
  T.~Appelquist, M.~Soldate, T.~Takeuchi and L.~C.~R.~Wijewardhana, 
 {\it ibid.}. 
 
%
 \bibitem{RGNJL}  K.-I.~Kondo, M.~Tanabashi, and K.~Yamawaki, Prog. 
  Theor. Phys. {\bf 89} (1993) 1249; M.~Harada, Y.~Kikukawa, T.~Kugo, 
  and H.~Nakano, Prog. Theor. Phys. {\bf 92} (1994) 1161. 

 \bibitem{MTY} V.~A.~Miransky, M.~Tanabashi and K.~Yamawaki, 
  Phys. Lett. {\bf B221}(1989)177 ; Mod. Phys. Lett. {\bf A4}(1989)1043;
  Y.~Nambu, Enrico Fermi Institute Report No. 89-08, 1989 
  (unpublished); in {\it Proceedings of the 1989 Workshop 
   on Dynamical Symmetry Breaking}, edited by T. Muta and K. Yamawaki 
  (Nagoya University, Nagoya, Japan, 1990);
  W.~A.~Bardeen, C.~T.~Hill, and M.~Lindner, Phys. Rev. {\bf D41}(1990)1647.

%

 \bibitem{STY} S.~Shuto, M.~Tanabashi and K.~Yamawaki, in 
              {\it Proceedings of the 1989 Workshop on 
              Dynamical Symmetry Breaking, Nagoya, Dec. 21-23, 1989}, 
             edited by T. Muta and K. Yamawaki 
                     (Nagoya University, Nagoya, Japan, 1990). 

 \bibitem{PCDC} V.~P.~Gusynin and V.~A.~Miransky, Phys. Lett. {\bf B198}
                (1987) 79. 

 \bibitem{CJT} J.~M.~Cornwall, R.~Jackiw, and E.~Tomboulis, 
  Phys. Rev. {\bf D10} (1974) 2428; 
  For a recent review, R.~W.~Haymaker, Riv. Nuovo Cim. {\bf 14} (1991) 1.

 \bibitem{KSY} V.~A.~Miransky and K.~Yamawaki, Mod. Phys. Lett. 
  {\bf A4}(1989)129. 

 \bibitem{cw} M.~Carena and C.~E.~M.~Wagner, Phys. Lett. {\bf B285} (1992) 277.

 \bibitem{bifurcation} D.~Atkinson, J. Math. Phys. {\bf 28} (1987) 2494.    

%
%

 \bibitem{Higgsmass_RGE} W.~A.~Bardeen and S.~T.~Love, Preprint 
	FERMILAB-PUB-91/233-T. 

 \bibitem{Higgsmass_WT} Shen~Kun and Q.~Zhongping, Phys. Rev. {\bf D45} 
  (1992) 3877; K.~Shen and Y.~P.~Kuang, Phys. Rev. {\bf D57} (1998) 6386. 
 
%
%
%

 \bibitem{DSBbook} See, for example, 
  V.~A.~Miransky, {\it Dynamical Symmetry Breaking 
  in Quantum Field Theories} (World Scientific Pub. Co. , Singapore 1993).

 \bibitem{ATW} T.~Appelquist, J.~Terning and 
  L.~C.~R.~Wijewardhana, Phys. Rev. {\bf D44} (1991) 871.    

 \bibitem{Gusynin} V.~P.~Gusynin and M.~Reenders, Phys. Rev. {\bf D57} (1998)
  6356. 

 \bibitem{PS} H.~Pagels and S.~Stokar, Phys. Rev. {\bf D20} (1979) 2947. 

%
%
%
 \bibitem{higashijima} V.~A.~Miransky, Sov. J. Nucl. Phys. {\bf 38} 
  (1983) 280; K.~Higashijima, Phys. Rev. {\bf D29} (1984) 1228.    

 \bibitem{pdg} Particle Data Group, Phys. Rev. {\bf D54} (1996) 1.    

 \bibitem{sigmameson} M.~Harada, F.~Sannino and J.~Schechter, 
  Phys. Rev. {\bf D54} (1996) 1991; S.~Ishida, M.~Ishida, 
  H.~Takahashi, T.~Ishida, K.~Takamatsu and T.~Tsuru, 
  Prog. Theor. Phys. {\bf 95} (1996) 745; S.~Ishida, T.~Ishida, 
  M.~Ishida, K.~Takamatsu and T.~Tsuru, 
  Prog. Theor. Phys. {\bf 98} (1997) 1005; M.~D.~Scadron, 
  hep-ph/9710317; V.~Elias, A.~H.~Fariborz, F.~Shi and T.~G.~Steele, 
  Nucl. Phys. {\bf A633} (1998) 279. 
 
 \bibitem{kugomich} K.-I.~Aoki, M.~Bando, T.~Kugo, M.~G.~Mitchard and 
  H.~Nakatani, Prog. Theor. Phys. {\bf 84} (1990) 683; 
  T.~Kugo and M.~G.~Mitchard, Phys. Lett. {\bf B286} (1992) 355. 
\end{thebibliography}
\end{document}